# Optical Resonators in Current and Future Experiments of the ALPS Collaboration

T. Meier[a] for the ALPS collaboration

[a]*Max-Planck-Institute for Gravitational Physics, Albert-Einstein-Institute, and Institut für Gravitationsphysik, Leibniz Universität Hannover, Callinstr. 38, D-30167 Hannover, Germany*

**Abstract.** The ALPS collaboration runs a "light shining through a wall" (LSW) experiment to search for weakly interacting sub-eV particles (WISPs). Its sensitivity is significantly enhanced by the incorporation of a large-scale production resonator and a small-scale high-power resonant second harmonic generator. Here we report on important experimental details and limitations of these resonators and derive recommendations for further experiments. A very promising improvement for a future ALPS experiment is the incorporation of an additional large-scale regeneration resonator. We present a rough sketch of how to combine a regeneration resonator with a single-photon counter (SPC) as detector for regenerated photons.

## INTRODUCTION

Although high-energy collider experiments are steadily optimized for searches for new particles with higher and higher masses, nothing forbids that yet unobserved particles may also exist at very low energy scales in the meV region. In this case their interaction with ordinary matter would be very weak [1]. Detection of such weakly interacting sub-eV particles (WISPs) needs a different type of experiment [2], that is optimized for high primary flux, low event rates and special detection schemes not based on the interaction of a WISP with matter.

Possible masses and coupling constants of WISPs are constraint by several experiments and astrophysical observations but especially the mass range from $10^{-6}$ eV to $10^{-2}$ eV lacks reliable experimental results [1-3]. Additionally, constraints from searches for WISPs of astrophysical origin can only rely on assumptions on the strength of the particle source because it is not under the control of the observer.

Hence the ALPS collaboration runs an LSW experiment to search for such types of WISPs that couple to a light and/or magnetic field, namely the Peccei-Quinn axion [4], other axion-like particles [1,5], massive hidden sector photons [2,5] and mini-charged particles [3,5]. The basics of LSW experiments are extensively explained in [5].

If these WISPs exist they could be the solution for several puzzling problems like the strong CP problem [1] or observations of anomalous transparency for high energy radiation in the universe [6-9]. They are also good cold dark matter candidates [1,10] and give hints on the validity of string theory models [2].

# ALPS – ENHANCING AN LSW EXPERIMENT WITH CAVITIES

There are several LSW experiments all over the world, for instance [11-13]. The sensitivity of those experiments depend mainly on four parameters, the magnetic field strength $B$, its length $L$, the flux of primary photons $N_{ph}$ and the dark count rate of the detector $N_d$ [5]. However, $B$ and $L$ are rather fixed by the available cryogenic magnet technology. In all experiments listed above pulsed lasers were used to provide $N_{ph}$. As such lasers are scaled up in output power their beam profile gets distorted from thermal and nonlinear effects inside their cavities. This requires larger apertures for magnets and detectors. While the magnet apertures are limited to approximately 4 cm by current designs, larger detector apertures often tend to increase $N_d$. Thus this way of enhancing the experiment's sensitivity appears to be somewhat limited.

In contrast, the ALPS collaboration pursues a different strategy. In our experiment at DESY in Hamburg, Germany we combine a large-scale optical resonator (or cavity) encompassing the production region and a high power infrared continuous-wave laser. The latter was provided by the Laser Zentrum Hannover (LZH), Germany. The optical resonator recycles most of the laser photons after each trip through the magnet, which can be understood in analogy to its use as power recycling cavity in gravitational wave detectors [14]. The coherent power enhancement effect inside the cavity, called power build-up, relaxes the amount of power needed from the primary laser for the same value of $N_{ph}$. Simultaneously the cavity serves as a spatial mode filter for the incident laser light. This keeps the required aperture of magnet and detector as small as possible supporting efforts to increase magnet length and achieve low dark count rates. The basics of optical resonators are described in [15]. To convert the incident laser light to the detector's spectral sensitivity maximum at 532 nm we introduced a nonlinear frequency doubling PPKTP crystal into the infrared laser beam [16].

At the same time incorporation of such large scale optical resonators with lengths of approximately 9 m significantly increase the complexity of every LSW experiment. Thus we first conducted a proof of principle experiment to show the feasibility of this technique. This was done successfully and the results are extensively discussed in [5].

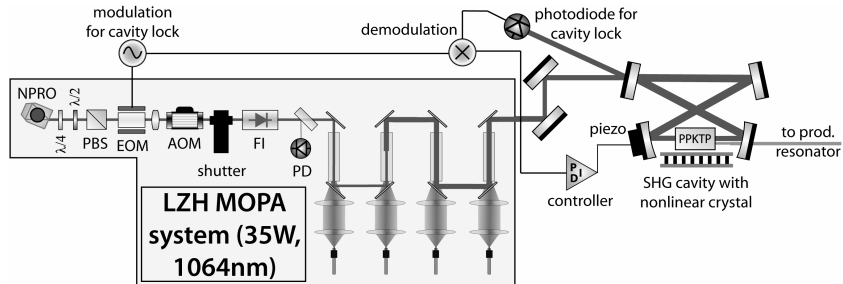

**FIGURE 1.** The improved optical setup of the resonant second harmonic generator (SHG). For details regarding the LZH MOPA system see references in [5].

In the next step we used the experience gained from our proof of principle experiment to enhance the sensitivity of the complete ALPS setup as far as possible given that magnet and location could not be changed. First, the amount of incident green light was increased by incorporation of a second optical resonator into the system that encompasses the nonlinear crystal. The optical setup is shown in Fig. 1.

As can be seen in Fig. 1 the resonator is designed as a folded ring resonator. This design makes sure that all converted light is emitted towards the production resonator. It is kept resonant with the incident laser light by an electronic feedback loop acting on its cavity length. The laser frequency stabilization to the production resonator and the length stabilization of the SHG resonator work simultaneously. Figure 2a shows the emitted power levels at 532 nm.

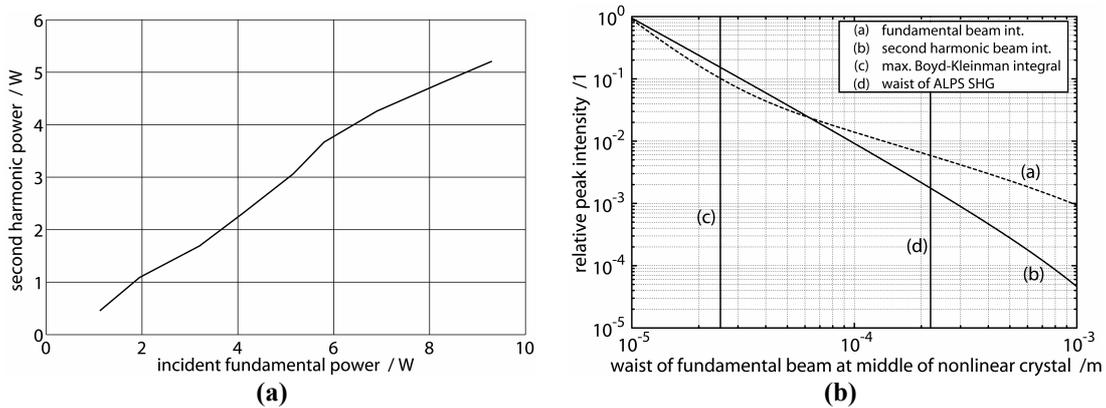

**FIGURE 2.** Part **(a)** shows the power at 532 nm that was emitted by the resonant second harmonic generator (SHG). Part **(b)** shows a simulation of the fundamental and second harmonic intensities at the position of the fundamental waist in the middle of the nonlinear crystal. Both intensities were scaled by different arbitrary values. Parameters were chosen as in the ALPS SHG with roundtrip losses of 1 %, an effective nonlinearity $d_{eff}$=7.5 pm/V and an input power of 8.0 W. It is assumed that impedance matching is re-established for each point on the x-axis by insertion of a new coupling mirror with suitable transmission. The vertical lines mark specific waist sizes discussed in the text.

After a few days of operation its efficiency slightly degraded. In its working point for long-term operation the SHG emitted 5 W at a wavelength of 532 nm from an incident power of 10 W at 1064 nm. The basics of second harmonic generation are extensively discussed in [17]. It is not straight-forward to achieve such power levels of 532 nm light efficiently because of deteriorating effects inside the nonlinear crystal like nonlinear absorption processes, gray-tracking and thermal dephasing. All these effects depend strongly on the intensity inside the crystal of either the fundamental beam or the second harmonic beam or both [18,19]. Figure 2b shows the dependence of both intensities on the size of the fundamental waist inside the crystal. For the 2 cm long PPKTP crystal that is used for the ALPS SHG the conventionally chosen fundamental waist size that maximizes the value of the Boyd-Kleinman integral is approximately 25 µm [17]. Instead we chose a 9 times bigger waist resulting in a peak intensity inside the nonlinear crystal that is 15 times lower for the fundamental and 100 times lower for the second harmonic beam. The waist of the second harmonic beam is always a factor of $\sqrt{2}$ smaller than the fundamental waist [17], forcing a quadratic reduction of the second harmonic intensity with increasing fundamental waist size. The slope of curve (b) in Fig. 2b does not deviate much from this showing that the available second harmonic output power is not harmed by our chosen enlargement factor of the fundamental waist as long as impedance matching of the SHG resonator to the incident fundamental beam is always re-established. Thus we reduced the above mentioned deteriorating effects and still achieved a long-term constant output power of 5 W. But even with this large waist size of 220 μm green output powers above approximately 8 W caused gray-tracking in this type of crystal. Unfortunately, the beam radius can not be enlarged significantly further because PPKTP crystals with heights bigger than 1 mm have not been available.

As second improvement we increased the power build-up of the ALPS production cavity. We placed both cavity mirrors into the production vacuum chamber to minimize losses for the circulating light by getting rid of vacuum windows in between the resonator mirrors (see [5] for an explanation of that problem). Special mirror mounts were assembled that allow for remote alignment, are vacuum compatible and, in the case of the end mirror inside the magnet, non-magnetic and only 35 mm in diameter. By this improvement the internal losses of the production resonator were reduced by roughly an order of magnitude. The new optical setup of the production resonator is shown in Fig. 3a. The explained improvements led to an average power inside the production resonator of 1200 W corresponding to a flux of primary particles of $3 \times 10^{21}$ photons/s. The mid-term fluctuations of the primary flux over time (i.e. the fluctuations of the circulating power) are shown in Fig. 3b. Further improvements and the results of the corresponding WISP search will be reported in [20].

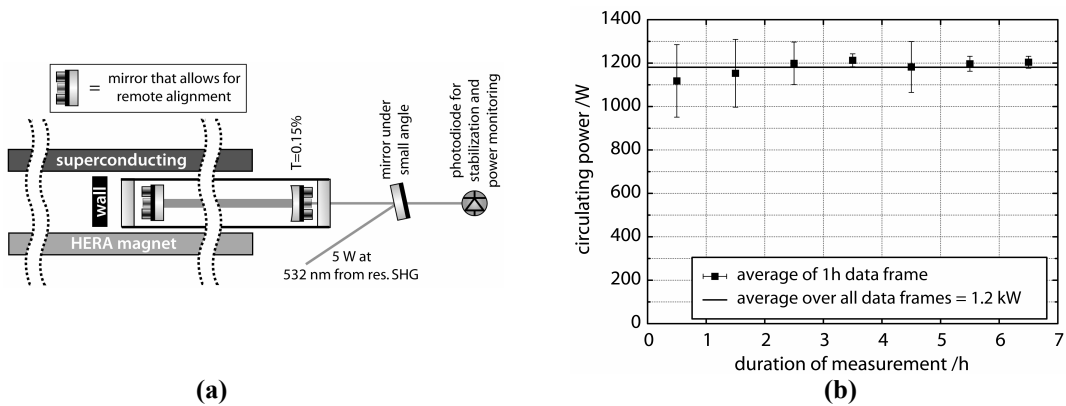

**FIGURE 3.** Part **(a)** shows the new experimental setup of the production resonator. Part **(b)** is an exemplary time series of the circulating power inside the production resonator. The data points are averages over 720 measurement points within 1 h of WISP search. The mean power is 1200 W.

To obtain amplification for the incident laser light it must be resonant inside a cavity. Due to the large length of the ALPS production resonator its resonance linewidth is reduced to 13 kHz. In an uncontrolled state length fluctuations of the optical resonator allow the laser light to be resonant for time segments of typically only 300 μs. This duration starts to approach the order of the cavity decay time, which will even be surpassed in future experiments with increased length and power build-up of the production resonator. Lock acquisition and control loop optimization will then become very difficult. Therefore one should think about efficient vibration isolation systems or quieter sites for such upcoming experiments.

The remaining losses inside the production cavity of 0.13 % per mirror originate most likely from the type of polish and coating used for the mirrors. We used mirrors with a planarity of λ/10 according to MIL-O-1380A that were coated by the so-called e-beam technique, which are relatively cheap and fast to obtain. In other experiments with comparable mirrors at a wavelength of 808 nm we measured losses per mirror by absorption and scattering of 0.14 %, which is in very good agreement with the value given above. Additionally, the ALPS production cavity mirrors showed a limited lifetime of approx. 25 hours at maximum power level. This effect is currently not understood but might originate from the combination of the chosen coating technique and laser wavelength of 532 nm. There are hints from older experiments in gravitational wave interferometry that coatings tend to degrade at these wavelengths on mid-term timescales [21]. Therefore to improve the experiment towards higher circulating power the laser wavelength should be changed to 1064 nm and mirrors should be tested that are polished and coated with more elaborate techniques like superpolishing and ion beam sputtering (IBS). Such optics are used and proven in the field of gravitational wave interferometry. To maintain their very low losses during the setup and operation of an improved experiment they should be used inside low-class cleanroom facilities only.

# RESONANT REGENERATION WITH A SINGLE PHOTON COUNTER

For some types of WISPs possible sources of astrophysical origin produce constraints on their properties surpassing those set by the current ALPS experiment (for instance the CAST experiment [22]). But in these experiments the strength of the source of the particles searched for is not under the experimentalist's control. This is different in LSW experiments. To achieve a sensitivity comparable to [22] with ALPS, one has to implement, besides other improvements, a new technique, namely a regeneration cavity with very high power build-up.

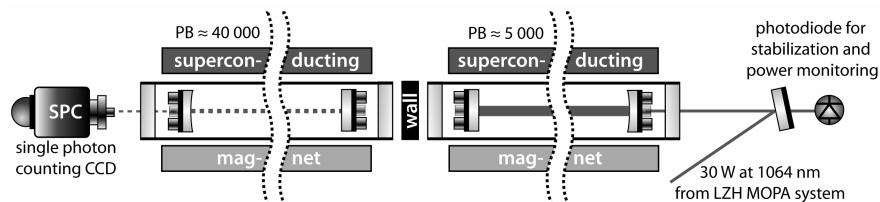

**FIGURE 4.** Schematic of production and regeneration resonator of a possible future ALPS experiment at a wavelength of 1064 nm. The values of input power and power build-ups (PB) give a realistic example for the use of cavity mirrors fabricated with currently best polishing and coating technologies

The regeneration cavity is yet another large-scale optical resonator encompassing the regeneration region [23]. It amplifies the regenerated electric field by coherent superposition in the same way as signal recycling cavities do in gravitational-wave detectors [14]. In LSW experiments its power build-up can be made very large because it is empty and the circulating power inside the regeneration resonator is tiny. Figure 4 shows a possible schematical layout of an LSW experiment at a laser wavelength of 1064 nm that combines production and regeneration resonator and still uses a single-photon counter (SPC). Further use of our present SPC is desirable because it is handy, easy to use and very well characterized. Such a setup is currently under investigation for a future ALPS upgrade.

For a resonant regeneration experiment to work properly the resonance frequency and alignment of the regeneration cavity mode must match the production cavity mode. Furthermore the sensitivity of the SPC must not be harmed by the light used for this stabilization. We propose to use frequency-doubled light to stabilize the regeneration resonator and then use dichroic optics to shield the SPC from it. Figure 5 shows a proposed setup.

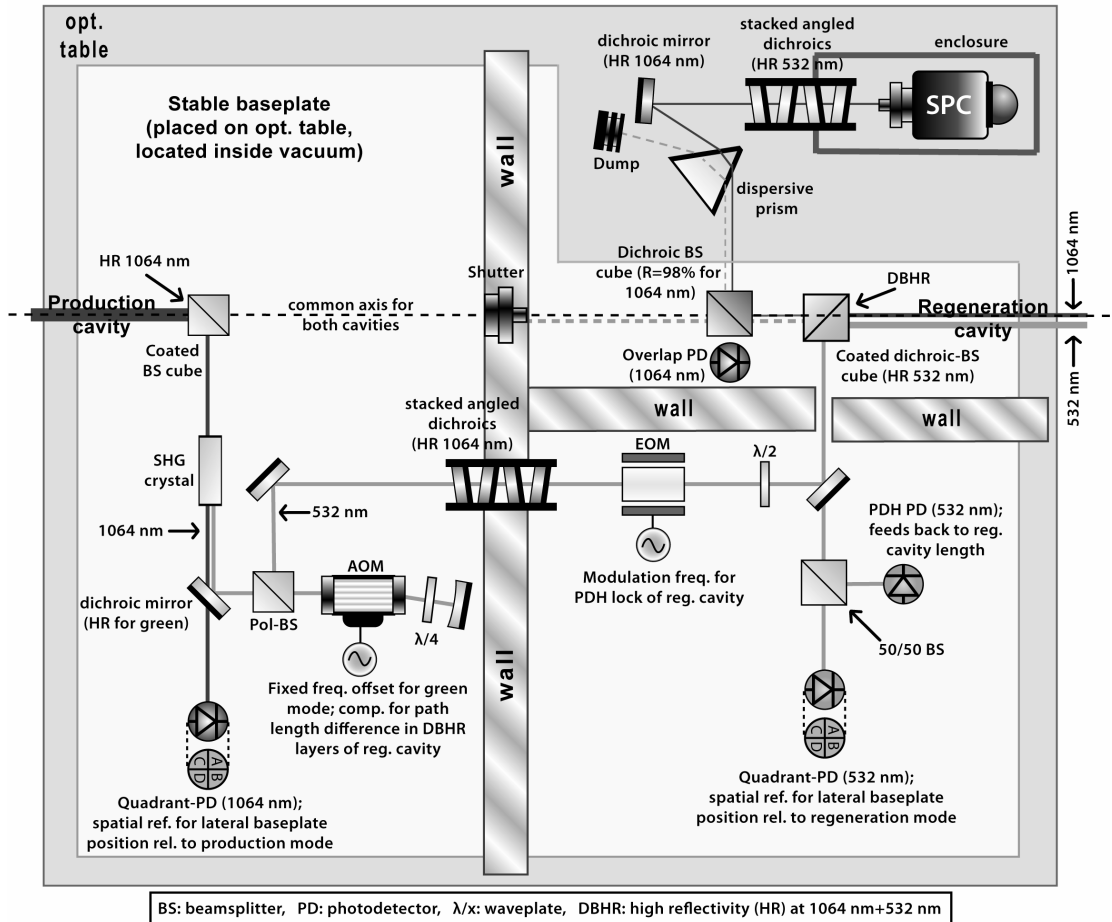

**FIGURE 5.** Optical setup to stabilize resonance and alignment of the regeneration cavity to the production cavity. Stabilization is done with frequency-doubled light at 532 nm.

The baseplate will be placed in between the pair of cavities. The coated beam splitter cubes act as plane inner end mirrors of the two cavities. They have to be aligned carefully such that their coated surfaces will be plane parallel. This assures that both cavity modes are oriented parallel. Two quadrant photodetectors will be used to fix their lateral position on the end mirror facets forcing them to share the same axis. The baseplate material and mounting procedure of the optics will be chosen in a way that no significant relative movement of the optics will happen.

A single-pass SHG will generate a few microwatts of green light from a small amount of transmitted light from the production resonator. Because the frequency is exactly doubled in this process the green light will share all resonances of the fundamental light inside the regeneration cavity. Correspondingly it will be used to stabilize the regeneration resonance. The acusto-optic modulator is used to compensate for possible different penetration depths of light with different wavelengths into the DBHR coatings of the regeneration cavity (see bottom of Fig. 5). Stacked and angled dichroic optics and dispersive elements will shield the single-photon counter from all light that was not regenerated from WISPs behind the wall.

A more detailed description of this resonant regeneration scheme will be subject of an upcoming publication [24].

## ACKNOWLEDGMENTS

We acknowledge financial support from the Helmholtz Association and from the Centre for Quantum Engineering and Space-Time Research (QUEST).